\documentclass[sigconf]{acmart}
\AtBeginDocument{%
  }

\usepackage{amsmath}
\usepackage{amsmath,amsfonts}
\usepackage{multirow}
\usepackage{subcaption}
\usepackage{graphicx}
\usepackage{bbm}
\usepackage[dvipsnames]{xcolor}

\usepackage{algorithm}
\usepackage{algorithmic}

\copyrightyear{2027}
\acmYear{2027}
\setcopyright{cc}
\setcctype{by}
\acmConference[WSDM '27]{Proceedings of the Twentieth ACM International Conference on Web Search and Data Mining}{February 15--19, 2027}{Hong Kong, Hong Kong}
\acmBooktitle{Proceedings of the Twentieth ACM International Conference on Web Search and Data Mining (WSDM '27), February 15--19, 2027, Hong Kong, Hong Kong}
\acmDOI{10.1145/3862322.3863718}
\acmISBN{979-8-4007-3117-4/2027/02}

\begin{document}


\title[GuardAgent]{GuardAgent: A Memory-Augmented Agentic Framework for Adaptive Detection of Adversarial Promotions in LLM-Based Recommendation}

\author{Shiyi Yang}
\orcid{0000-0002-0539-8391}
\affiliation{%
  \institution{The University of New South Wales}
  \city{Sydney}
  \state{NSW}
  \country{Australia}}
\affiliation{%
  \institution{CSIRO}
  \city{Sydney}
  \state{NSW}
  \country{Australia}}
\email{shiyi.yang1@unsw.edu.au}

\author{Xinshu Li}
\orcid{0000-0002-1202-3993}
\affiliation{%
  \institution{Monash University}
  \city{Melbourne}
  \state{VIC}
  \country{Australia}}
\affiliation{%
  \institution{Macquarie University}
  \city{Sydney}
  \state{NSW}
  \country{Australia}}
\email{xinshu.li@monash.edu}

\author{Chen Wang}
\orcid{0000-0002-3119-4763}
\affiliation{%
\institution{CSIRO}
\city{Sydney}
\state{NSW}
\country{Australia}}
\affiliation{%
  \institution{The University of New South Wales}
  \city{Sydney}
  \state{NSW}
  \country{Australia}}
\email{chen.wang@csiro.au}

\author{Xiwei Xu}
\orcid{0000-0002-2273-1862}
\affiliation{%
\institution{CSIRO}
\city{Sydney}
\state{NSW}
\country{Australia}}
\affiliation{%
  \institution{The University of New South Wales}
  \city{Sydney}
  \state{NSW}
  \country{Australia}}
\email{xiwei.xu@csiro.au}

\author{Liming Zhu}
\orcid{0000-0001-5839-3765}
\affiliation{%
\institution{CSIRO}
\city{Sydney}
\state{NSW}
\country{Australia}}
\affiliation{%
  \institution{The University of New South Wales}
  \city{Sydney}
  \state{NSW}
  \country{Australia}}
\email{liming.zhu@csiro.au}

\author{Lina Yao}
\orcid{0000-0002-4149-839X}
\affiliation{%
  \institution{The University of New South Wales}
  \city{Sydney}
  \state{NSW}
  \country{Australia}}
\email{lina.yao@unsw.edu.au}


\renewcommand{\shortauthors}{Shiyi Yang et al.}

\begin{abstract}


In traditional recommender systems (RSs), textual information typically serves as auxiliary signals for representation learning. In contrast, in LLM-based RSs, textual information becomes the primary medium for recommendation generation through semantic understanding and reasoning.
This shift fundamentally changes the attack surface, making promotion-driven manipulations more feasible, effective, and harder to detect. However, effectively detecting such attacks remains largely unexplored. 
To fill this gap, we present the first systematic study on detecting promotion-driven attacks against LLM-based RSs, aiming to safeguard next-generation recommendation platforms. Specifically, we propose GuardAgent, a novel LLM-powered agentic detection framework that operates through a loop of Perception, Action, Reflection and Memory. In contrast to conventional LLM-based detectors that function in a static, one-off, and opaque fashion, GuardAgent continually learns and self-evolves to adapt to unseen threat patterns while generating explainable reasoning traces and outcomes. Extensive experiments on real-world datasets demonstrate that GuardAgent achieves state-of-the-art generalization performance, laying the foundation for robust and trustworthy recommendation in practical settings.


\end{abstract}



\maketitle

\section{Introduction} \label{intro}

Recommender systems (RSs) are a vital component of modern digital platforms, enabling users to discover desirable products while boosting providers’ revenue through targeted item promotion \cite{ge2024survey}. Over time, RSs have evolved from relying solely on user–item interactions to incorporating auxiliary content signals such as product images and titles, aiming to alleviate issues such as data sparsity and cold start \cite{yang2024attacking, zhang2024stealthy}. Such multimodal features serve as complementary signals to enhance representation learning \cite{zhang2019deep}. In contrast, the emergence of large language model (LLM)-based RSs fundamentally shifts this paradigm. Recent models \cite{li2023text, geng2022recommendation, bao2023tallrec, zhang2025collm, zhang2024agentcf} generally formulate recommendation as a language-driven task, where user behavior, item information, and task objectives are expressed in natural language prompts, and personalized recommendations are produced through language understanding, reasoning, and generation \cite{zhao2024recommender}. As a result, textual content is no longer a side feature, but becomes the primary interface through which the model performs recommendation decisions \cite{zhao2024recommender, huang2025survey}. However, this paradigm shift also broadens the attack surface, raising new security and robustness challenges \cite{zhang2024stealthy, ning2024cheatagent, yang2026drunkagent, chua2024ai}.



\begin{figure}[t]
    \centering
     \includegraphics[width=0.9
     \linewidth]{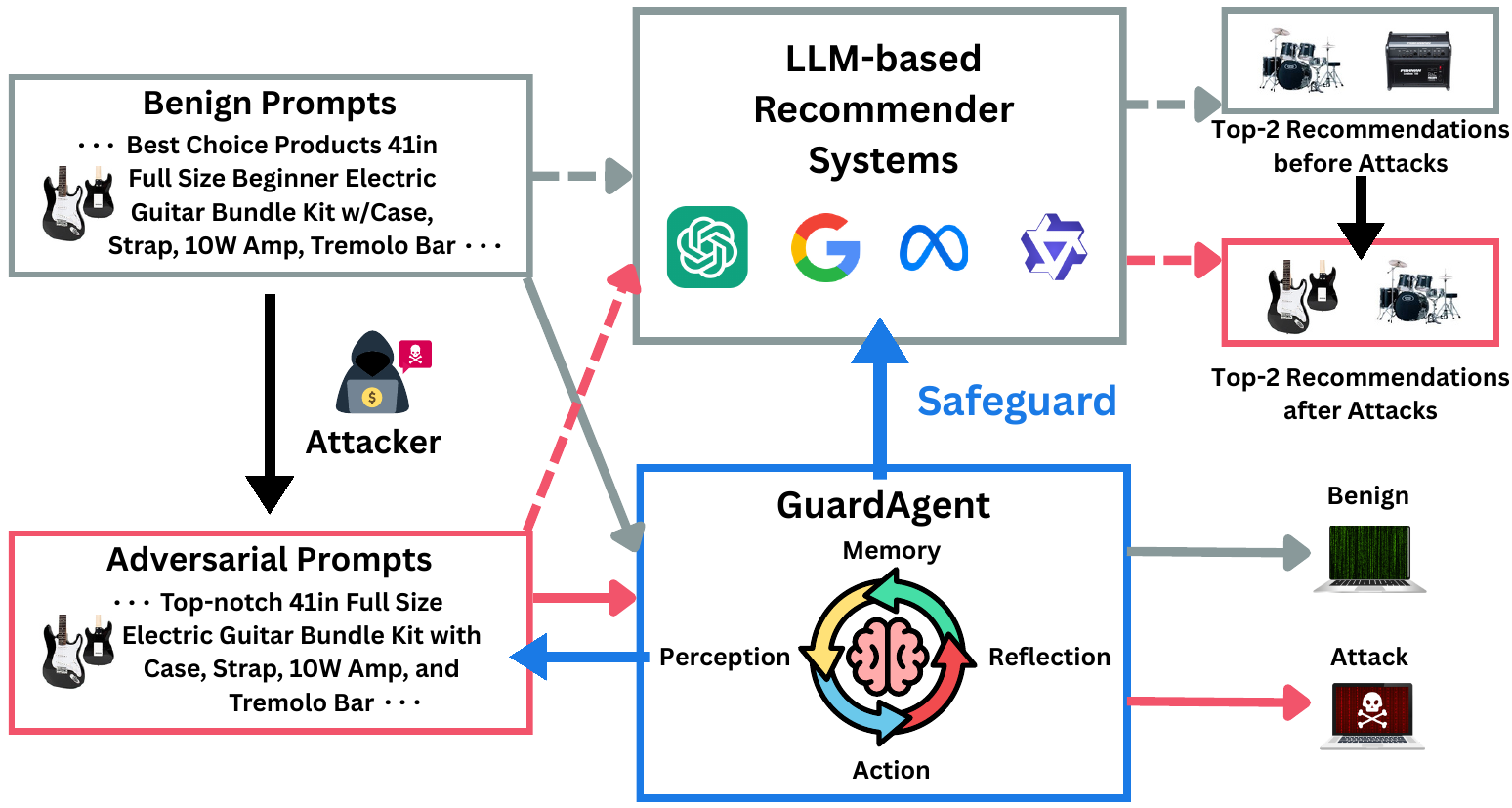}
     \vspace{-1em}
    \caption{GuardAgent proactively detects prompt perturbations in LLM-based recommenders, contributing to the trustworthiness of next-generation recommendation platforms.}
    \label{fig:intro_fig}
    \vspace{-2.6em}
\end{figure}

RSs have long been recognized as highly vulnerable to adversarial manipulation, particularly item promotion attacks that aim to increase the exposure of target items \cite{feng2025attacks, wang2026secure}, as illustrated in Fig.~\ref{fig:intro_fig}, thereby fostering unfair competition among providers and eroding users’ trust in the platform ecosystem. Such attacks align closely with real-world threats observed on platforms like Amazon, eBay, and Sony \cite{lin2020attacking, lin2022shilling, yang2026drunkagent}. Existing attacks are predominantly realized through data poisoning, where malicious actors, typically profit-driven item providers, inject well-crafted structured interactions (e.g., ratings or purchases) into the training process of RSs to promote their target items \cite{nguyen2024manipulating, wang2024poisoning, gu2026llm}. In response, existing defenses focus mainly on identifying and filtering such anomalous interactions \cite{wang2024poisoning, wang2023recad, feng2026there}. However, current attacks against LLM-based RSs often operate at inference time through prompt perturbations in the open-ended language space \cite{zhang2024stealthy, ning2024cheatagent, yang2026drunkagent}, as shown by the example in Fig.~\ref{fig:intro_fig}, limiting the effectiveness of those existing defenses. Consequently, systematic detection methods tailored to LLM-based RSs remain largely underexplored.

Existing detection approaches for LLMs (e.g., perplexity-based detectors), which are primarily designed to recognize adversarial prompts that induce unsafe or misaligned outputs \cite{liu2024formalizing, zhang2025agent}, are not well-suited to the characteristics of RS attacks. Moreover, these methods typically operate in a static and one-off manner, lacking mechanisms to adapt to evolving attack strategies \cite{wang2026secure}. This limitation becomes more pronounced as recent attacks increasingly leverage LLMs to generate more effective, human-like, and stealthy promotional content \cite{zhang2024stealthy, yang2026drunkagent}. As a result, these detectors often exhibit degenerate behaviors, either overwhelmingly favoring benign predictions or achieving high detection rates at the cost of excessive false alarms (as demonstrated in Section~\ref{exp}). Beyond detection performance, explainability is essential for real-world deployment \cite{li2023survey}. However, most existing detectors \cite{wang2023recad, wang2024poisoning} operate as black boxes and provide detection decisions without exposing a human-understandable decision process. This highlights the urgent need for adaptive and explainable detectors for LLM-based RSs.







To address the above challenges, in this paper, we propose GuardAgent, a novel LLM-powered agentic detection framework designed to safeguard LLM-based RSs against promotion-driven adversarial content manipulations. We  formulate the framework as a cognitive agentic loop consisting of Perception, Action, Reflection, and Memory, as illustrated in Fig.~\ref{fig:intro_fig}. Specifically, GuardAgent maintains a dynamic memory mechanism, allowing it to continually interact with and learn from its environment, and to evolve its decision policies through self-reflection on past detection failures and erroneous reasoning traces, thereby adapting to previously unseen benign and attack patterns. To improve the explainability, GuardAgent distills reasoning into memory-augmented decision policies and explains its detection outcomes via explicit policies and reasoning traces, enabling security teams to audit decisions, diagnose potential failures, and respond effectively to emerging threats. 




Our main contributions are summarized as follows:

\begin{itemize}

\item To the best of our knowledge, we present the first systematic study on detecting adversarial attacks with the aim of promoting target items in LLM-based RSs, for safeguarding the next generation RSs from real-world threats.

\item We propose to formulate adversarial detection as a cognitive agentic loop, enabling continual learning and self-evolution to adapt to dynamic environments where previously unseen threats constantly emerge.

\item We propose GuardAgent to recognize adversarial content manipulations in recommendations and provide explainable detection outcomes through explicit textual policies and reasoning traces.

\item Extensive experiments on real-world datasets show that GuardAgent consistently outperforms state-of-the-art baselines, exhibiting strong generalization performance.

\end{itemize}

\begin{figure*}[t]
    \centering
    \includegraphics[width=0.8\linewidth]{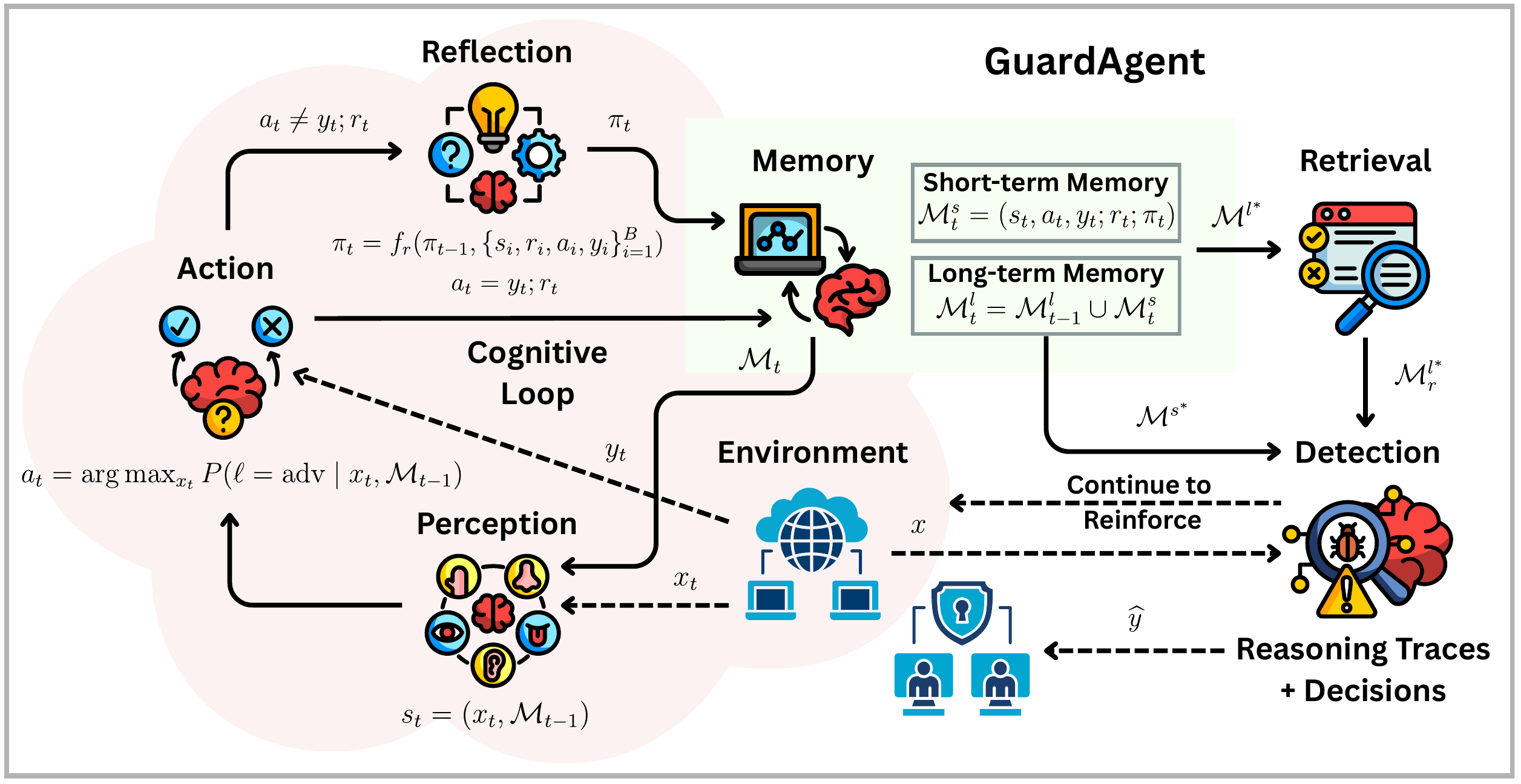}
    \vspace{-1.4em}
      \caption{GuardAgent Overview. GuardAgent is designed as a cognitive loop that iteratively performs Perception, Action, Reflection, and Memory operations, enabling the agent to continuously learn and self-evolve, thereby adapting to dynamic recommendation environments where new threats constantly emerge. Through this loop, GuardAgent  enhances robustness against evolving adversarial perturbations and produces explainable and trustworthy detection results.}
    \label{fig:system_overview}
    \vspace{-1.2em}
\end{figure*}

\section{Problem Formulation}

\subsection{Threat Model} \label{threat_model}

\subsubsection{\textbf{Attacker’s  Knowledge:}} \label{attack_knowledge}

In real-world recommender systems, it is generally infeasible for adversaries to access the internal details of the victim LLM-based RSs \cite{li2023text, geng2022recommendation, bao2023tallrec, zhang2025collm}, such as parameters and architectures, as highlighted in previous works \cite{ning2024cheatagent, zhang2024stealthy, yang2026drunkagent}. Accordingly, we consider a black-box threat model for evaluating GuardAgent to align with the practical constraints. Given the inherently open nature of recommendation platforms \cite{si2020shilling, 10415763, yang2025retrieval}, the attackers are allowed to exploit publicly available resources, including user interactions and item metadata, to design iterative probing strategies that improve attack efficacy.

\subsubsection{\textbf{Attacker’s Objectives:}}
Let $\mathcal{U}$ and $\mathcal{V}$ represent the set of real users and the item universe, respectively, and $\mathcal{D} = \{r_{u,v}: u \in \mathcal{U}, v \in \mathcal{V}\}$ to denote the records of the user-item interaction matrix in the large-scale recommendation space, where $r_{u, v}$ indicates the interaction between the user $u$ and the item $v$. As discussed in Section~\ref{intro}, our primary attack objective is to maximize the hit ratio $h(v_t)$ of a target item $v_t$ on the black-box RSs, which aligns with the main security concerns faced in both practical deployments and academic studies \cite{lin2020attacking, lin2022shilling, yang2026drunkagent}. Formally, the ratio is the percentage of normal users whose recommendation lists include $v_t$ after the attack: $h(v_t) = \sum_{u\in \widehat{\mathcal{U}}_{v_t}} \frac{\mathbb{I}(v_t \in \widehat{ \mathcal{V}}_{u,:k})}{|\widehat{\mathcal{U}}_{v_t}|}$,  where $\mathbb{I}(\cdot)$ is an indicator function, $\widehat{\mathcal{U}}_{v_t} = \{u \in \mathcal{U}: r_{u,v_t} = 0\}$ is the set of normal users and $\widehat{\mathcal{V}}_{u,:k}$ is the top-$k$ recommendation list of the normal user $u$. Given that the practical attacks often have invisibility to maximize the attack impact \cite{zhang2024stealthy, yang2026drunkagent}, the secondary attack goal is to make the attack as stealthy as possible.

\subsubsection{\textbf{Attacker’s Capabilities:}} \label{attacker_cap}

In realistic settings, merchants are typically responsible for maintaining item metadata (e.g., item titles and item descriptions) through platform-provided APIs \cite{cohen2021black, yang2026drunkagent}. Since such textual metadata are embedded in prompts and directly participate in language understanding and generation processes in recommendation \cite{li2023text, geng2022recommendation, bao2023tallrec, zhang2025collm, zhang2024agentcf}, existing adversarial attack studies mainly assume that adversaries (usually profit-driven merchants) can manipulate these textual inputs to promote their target items in recommendation results \cite{zhang2024stealthy}. Accordingly, the adversary operates exclusively at the input level and can only interact with the victim system through queries and publicly exposed recommendation outputs to guide targeted adversarial perturbations.

\subsection{Defensive Mechanism}
We formulate adversarial content detection as an iterative sequential decision-making problem within an agentic framework. At each time step $t$, the agent perceives the detection environment and constructs a detection state $s_t=(x_t, \mathcal{M}_{t-1})$, where $x_t$ denotes the newly arrived content, and $\mathcal{M}_{t-1}$ represents the memory module containing decision policies accumulated from past interactions and reflections. Given this state, the agent takes an action $a_t$ by reasoning over its current memory: $a_t = f_a(s_t)$, where $f_a(\cdot)$ denotes the decision function. After receiving the ground-truth feedback $y_t$ from the environment, if the action is correct, this episode ends with success; otherwise, the agent conducts reflection, where it replays the decision-making context, analyzes its erroneous reasoning process, and derives refined textual policies $\pi_{t}$. These policies are then integrated into memory, producing the new memory $\mathcal{M}_{t}$, which can support more robust future reasoning and actions. As a result, the next state $s_{t+1}=(x_{t+1}, \mathcal{M}_{t})$ is formed. The overall objective is to maximize cumulative detection performance: 
\begin{equation} \label{eq2}
\max_{\mathcal{M}} \; \mathbbm{E}\!\left[\sum_{t=1}^{T} \mathbbm{I} {\,(a_t=y_t)\,}\right],
\end{equation} where $\mathbbm{I}(\cdot)$ is the indicator function  and $T$ denotes the total number of interactions. After sufficient iterations, the agent obtains an evolved memory state $\mathcal{M}^*$, which serves as a critical component for effective detection tasks. This iterative loop enables the agent to self-evolve, progressively enhancing its robustness and adaptivity in dynamic recommendation environments where unseen adversarial threats continually emerge. We detail the loop in the next section.


\section{Methodology}

In this section, we present our novel detection framework, GuardAgent. As shown in Fig.~\ref{fig:system_overview}, our framework is designed as a cognitive agentic loop that iteratively performs \textit{Perception}, \textit{Action}, \textit{Reflection}, and \textit{Memory}, enabling the agent to continuously self-evolve in a dynamic environment. The loop culminates in a \textit{Detection} phase, where the accumulated memory are deployed to recognize adversarial content in unseen data. This design ensures adaptivity during training and robustness as well as explainability during inference.

\subsection{Perception}

In each iteration $t$, GuardAgent interacts with the environment by perceiving newly arrived content data $x_t$. The input is then integrated with the agent’s memory $\mathcal{M}_{t-1}$ to form the detection state $s_t=(x_t, \mathcal{M}_{t-1})$, which unifies external environmental signals with accumulated internal knowledge, enabling the agent to act based on both current context and past experiences. By grounding every subsequent decision on this enriched state, the agent not only preserves persistence across interactions but also provides a robust basis for adaptive decision-making, continual learning, and ultimately robust malicious content detection.

\subsection{Action}
In this stage, GuardAgent reasons over the detection state $s_t=(x_t, \mathcal{M}_{t-1})$ to identify which candidate is malicious. Here, we instantiate the input $x_t$ as a pair of candidate contents, $x_t = \{x_{t_b}, x_{t_m}\}$, where $x_{t_b}$ denotes the benign candidate and $x_{t_m}$ represents the adversarial one. Rather than the fixed heuristics-based detectors \cite{liu2024formalizing}, the agent relies on dynamic memory, which maintains adaptive decision policies shaped by previous interactions and past detection outcomes, to make decisions. As such, the decision-action function $f_a(s_t)$ estimates the adversarial probability of a candidate conditioned on the current memory, and the action $a_t$ is then defined as selecting the candidate with the highest adversarial probability:
\begin{equation} \label{eq3}
a_t = \arg\max_{x \in \{x_{t_b}, x_{t_m}\}} P(\ell=\text{adv} \mid x, \mathcal{M}_{t-1}),
\end{equation} where $\ell=\text{adv}$ denotes the malicious class and $\ell=\text{benign}$ denotes the normal class. Here, $P(\ell=\text{adv} \mid x, \mathcal{M}_{t-1})$ reflects how strongly a candidate’s features align with the adversarial patterns represented in the memory-encoded policies.

In the implementation (as elaborated in Fig.~\ref{fig:guardagent_prompts} of Appendix~\ref{imple_guardagent}), we craft structured prompt templates on reasoning LLMs to instantiate $f_a(\cdot)$. Specifically, we explicitly specify the agent role, threat model, task goals as well as instructions, and output format constraints, ensuring consistent and reproducible agentic behaviors. To further improve effectiveness, we introduce structured chain-of-thought \cite{wei2022chain} (CoT) reasoning into the prompts, which guides the agent to perform step-wise pattern identification (Step 1) and rule alignment (Step 2) prior to classification and justification (Step 3). As a result, the agent yields a detection decision supported by reasoning traces. However, the raw traces are often lengthy and noisy; therefore, a reasoning-enhanced explanation $r_t$ is generated to distill the essential decision rationale. These explicit reasoning traces enhance explainability and offer informative internal signals for the subsequent Reflection stage, where systematic detection errors are analyzed and decision policies are appropriately refined. 



As illustrated in Fig.~\ref{fig:system_overview}, GuardAgent then receives the ground-truth feedback $y_t$ from the environment and compares it with its chosen action $a_t$. If the action correctly matches the ground-truth malicious candidate, i.e., $a_t=y_t$, the current episode terminates successfully, and the agent proceeds to the memory stage directly. Otherwise, if $a_t \neq y_t$, a reflection process is triggered to analyze the erroneous reasoning trajectories for policy refinement.

\subsection{Reflection}

The reflection stage aims to refine symbolic (i.e., rule-based and human-interpretable) decision policies by analyzing reasoning errors arising from incorrect detections, rather than relying on reward-driven optimization as commonly adopted in reinforcement learning-based agents \cite{afsar2022reinforcement}, which may reduce explainability and controllability in security-sensitive environments. To this end, GuardAgent introduces \textit{batch-wise and contrastive mechanisms} to support robust and generalizable policy refinement.

\subsubsection{Batch-wise Learning} Unlike most episodic reflection methods \cite{shinn2023reflexion, zhang2024agentcf, yang2026drunkagent} that perform updates based on isolated failures, GuardAgent learns decision policies over a batch of failure cases accumulated across multiple interactive detection episodes. Each case consists of a benign--malicious content pair exhibiting false alarms and missed detections, together with the associated mistaken reasoning traces. The batch-wise reflection improves the stability and generalizability of policy updates by mitigating overfitting to individual failures and reducing sensitivity to noisy error cases.

\subsubsection{Two-level Contrastive Analysis} To capture both individual errors and systematic limitations, we adopt a two-level contrastive analysis paradigm for policy refinement: 1) \textit{Pairwise Contrast} is conducted within each benign--malicious failure pair. By jointly revisiting the detection state $s_t$, the predicted action $a_t$, and the associated reasoning traces $r_t$, reflection identifies misleading patterns (e.g., generic marketing language or standard item descriptions) in benign content that triggered false alarms, as well as overlooked cues in adversarial content that failed to indicate malicious intent based on $y_t$. This localized contrast highlights relative pattern differences responsible for incorrect decisions, enabling precise correction of decision boundaries. 2) \textit{Contrastive Aggregation} is performed across multiple failure pairs within the batch. Patterns that consistently appear in omitted adversarial content are strengthened, while patterns frequently observed in benign content are suppressed or qualified. Through this consolidation update, the robustness and generalizability to unseen attacks are enhanced.


\subsubsection{Policy Refinement} The agent finally distills the above analysis into refinement signals in the form of revised or newly derived decision policies, where revision is prioritized to avoid memory explosion and ensure stable adaptation. Formally, the refined decision-making policies $\pi_{t}$ obtained through in-context learning is
\begin{equation} \label{eq4}
\pi_{t} = f_r(\pi_{t-1}, \{s_i, r_i, a_i, y_i\}_{i=1}^{B}),
\end{equation} where $B$ is the number of failure cases in the reflection batch, and $f_r(\cdot)$ denotes the reflection function. The function is implemented via structured prompting to reasoning LLMs and more details are given in Appendix~\ref{imple_guardagent}, where the task-aware context is enhanced via a well-designed CoT reasoning (Step 1-3 in Fig.~\ref{fig:guardagent_prompts}) to support the overall batch-wise contrastive analysis over failure cases for policy refinement. The refined policies are subsequently provided to the Memory stage for integration with existing knowledge, enabling continual improvement in future detections.

\subsection{Memory}

\subsubsection{Memory Initialization}

Before entering the cognitive loop, GuardAgent initializes its memory $\mathcal{M}_0$ with a small set of pre-defined detection policies induced by prompting LLMs. To enhance their generality, the prompts explicitly incorporate attack goals, encouraging the model to synthesize decision policies that capture common promotion-oriented manipulation patterns, such as lexical noise, exaggerated claims, and clickbait urgency \cite{zhang2024stealthy}. These policies act as heuristic knowledge priors that reduce cold-start uncertainty, constrain the search space, and accelerate the adaptation process.


\subsubsection{Short-term vs. Long-term Memory} During the loop, GuardAgent maintains a memory module that integrates various abstraction levels of traceable knowledge derived from its detection experiences. Specifically, the memory consists of three complementary components: 1) episodic evidence, 2) explicit reasoning traces, and 3) textual decision policies, collectively supporting the distinction between benign and malicious content. The hierarchy ensures high-level policies are experience-grounded and auditable, enhancing accountability by enabling each policy to be traced back to its underlying traces and evidences, while simultaneously providing a basis for trustworthy adaptivity to previously unseen threats.

\begin{algorithm}[t]
\caption{GuardAgent: Memory-Augmented Reasoning Agents for Detecting Promotion Attacks in LLM-Based RSs}
\label{alg:guardagent}
\small
\begin{algorithmic}[1]

\REQUIRE
Interaction episodes $\{(x_t,y_t)\}_{t=1}^{T}$,
initial memory $\mathcal{M}_0$,
batch size $B$, incoming query samples $\{x\}$

\ENSURE
Optimized memories $\mathcal{M}^{s^*}$ and $\mathcal{M}^{l^*}$,
detection predictions $\{\widehat{y}\}$

\STATE Initialize short-term memory $\mathcal{M}_0^s$ and long-term memory $\mathcal{M}_0^l$
\STATE Initialize failure buffer $\mathcal{B}\leftarrow \emptyset$

\FOR[\textbf{Cognitive Loop}]{each interaction episode $(x_t, y_t)$}

    \STATE Construct detection state:
    $s_t=(x_t,\mathcal{M}_{t-1})$
    \COMMENT{\textbf{Perception}}

    \STATE Perform memory-augmented decision with Eq.~(\ref{eq3})
    \COMMENT{\textbf{Action}}

    \STATE Obtain $a_t$ and the reasoning-enhanced explanation $r_t$

    \IF{$a_t \neq y_t$}

        \STATE Store failure case $(s_t,r_t,a_t,y_t)$ into $\mathcal{B}$

    \ENDIF

    \IF[\textbf{Reflection}]{$|\mathcal{B}| \geq B$}

        \STATE Perform pairwise contrastive analysis within failure pairs

        \STATE Perform contrastive aggregation across failure cases

        \STATE Refine policies $\pi_{t}$ by analyzing reasoning errors with Eq.~(\ref{eq4})

        \STATE Clear failure buffer $\mathcal{B}\leftarrow \emptyset$

    \ENDIF

    \STATE Update short-term memory $\mathcal{M}_t^s$ with Eq.~(\ref{eq5})
    \COMMENT{\textbf{Memory}}

    \STATE Update long-term memory $\mathcal{M}_{t}^l$ with Eq.~(\ref{eq6})

\ENDFOR

\STATE Obtain optimized memories $\mathcal{M}^{s^*}$ and $\mathcal{M}^{l^*}$

\FOR[\textbf{Detection}]{each query sample $x$}

    \STATE Obtain the retrieved memory $\mathcal{M}_r^{l^*}$ with Eq.~(\ref{eq7})

    \STATE Perform retrieval-augmented policy detection on $x$ with Eq.~(\ref{eq8})

    \STATE Output prediction $\widehat{y}$ with human-readable reasoning traces

\ENDFOR

\end{algorithmic}
\end{algorithm}

To balance rapid adaptation and long-term robustness, GuardAgent builds memory along two timescales: short-term memory (STM) and long-term memory (LTM). STM provides an evolving repository that stores recently updated and interaction-level knowledge extracted from ongoing detections, allowing the agent to promptly respond to novel manipulations. Let $\mathcal{M}_{t}^s$ represent STM, \begin{equation} \label{eq5}
\mathcal{M}_{t}^s = (s_t, a_t, y_t; r_t; \pi_{t}).
\end{equation} In contrast, LTM (denoted as $\mathcal{M}_{t}^l$) serves as a persistent workspace that retains historically validated knowledge accumulated over time, providing continuity, protecting against catastrophic forgetting, and ensuring stability. Formally, the LTM consolidation is 
\begin{equation} \label{eq6}
\mathcal{M}_{t}^l = \mathcal{M}_{t-1}^l \cup \mathcal{M}_{t}^s,
\end{equation} where LTM is extended with the validated knowledge from STM. To sum up, STM’s fast adaptation and LTM’s durable retention, combined with the hierarchical organization, jointly form adaptive decision boundaries, making the reasoning process remain explainable and resilient to emerging adversarial strategies and tactics.



\subsection{Memory-Augmented Adversarial Detection}

After sufficient iterations, the agent obtains an evolved memory state $\mathcal{M}^*$ that enables reliable recognition of new adversarial content. To ensure the detection effectiveness, we incorporate the optimized STM $\mathcal{M}^{s^*}$, which captures the refined decision policies $\pi^*$ from recent interactions, with the LTM $\mathcal{M}^{l^*}$, which preserves accumulated precedents $\{s_j, r_j, a_j, y_j, \pi_j\}_{j=1}^{T}$ from historical experiences. To improve the threat adaptivity, GuardAgent retrieves the most relevant records of the current input $x$. Specifically, for each $x_{j}$, the similarity score $c_j$ between it and the query is computed as,
\begin{equation} \label{eq7}
c_j = f_s(x,x_j),
\end{equation} where $f_s(\cdot)$ is implemented via lightweight cosine similarity, i.e., $f_s(x,x_j) = \frac{f_e(x)\cdot f_e(x_j)}{||f_e(x)||||f_e(x_j)||}$, and $f_e(\cdot)$ is built with a pre-trained textual encoder (e.g., sentence transformer \cite{yang2026drunkagent}). As such, the indices of the top-$k$ most similar records are obtained as $
\mathcal{I}_k = \operatorname{argsort}_{j}(\{c_j\}_{j=1}^{T})_{1:k}$. Since some components (e.g., reasoning traces) are inherently time-dependent and reflect past decision contexts that may no longer be consistent with the current policies, they are not directly suitable for retrieval at inference time. Therefore, we rely on example–label pairs as stable retrieval units to provide concrete evidence to complement the abstract policies. As a result, the retrieved memory is $\mathcal{M}_r^{l^*} = \{(x_j, y_j)\mid j \in \mathcal{I}_k\}$. The prediction  $\widehat{y}$ is 
\begin{equation} \label{eq8}
\widehat{y} = f_d(x, \mathcal{M}^{s^*}, \mathcal{M}_r^{l^*}),
\end{equation} where $f_d(\cdot)$ represents a retrieval-augmented decision function to integrate the current input and the optimized and retrieved memories to yield a detection outcome. We instantiate $f_d$ using reasoning LLMs guided by carefully designed prompts enhanced via CoT reasoning for strengthening the robustness to unseen attacks, as illustrated in Fig.~\ref{fig:guardagent_prompts}. 
To bridge examples and policies, we generate reasoning traces from the examples that are explicitly aligned with the current policies. These traces enable analogy reasoning, allowing the agent to identify adversarial patterns and map them to relevant policies for accurate detection. Crucially, the final decisions are accompanied by faithful and human-readable reasoning traces that substantiate the agent’s judgments, providing trustworthy and actionable artifacts for security analysis and audit processes.

To sum up, GuardAgent operates in two stages: an adaptation stage and an inference stage. During the adaptation stage, the agent performs iterative reflection-based memory evolution within the cognition loop to refine its detection policies. During the inference stage, the evolved memory is retrieved to support retrieval-augmented reasoning for attack detection, as summarized in Algorithm~\ref{alg:guardagent}. As shown in Fig.~\ref{fig:system_overview}, detection outcomes serve as feedback signals to periodically reinforce the agent’s cognition, forming a continual process of improving resilience against evolving attacks.

\begin{table}
  \caption{Statistics of Datasets}
  \vspace{-1.5em}
  \begin{center}
  \small
  \label{tab:statistics}
  \resizebox{0.85\columnwidth}{!}{
 \begin{tabular}{c|cccc}
\hline
  \textbf{Dataset} & \textbf{\#Users} & \textbf{\#Items} & \textbf{\#Interactions} & \textbf{Sparsity}\\
\hline
  Amazon Products & 27,530 & 10,620 & 231,392 & 99.92\% \\

  Yelp Business & 1,599 & 1,318 & 30,120 & 98.57\% \\
\hline
  \end{tabular}}
  \end{center}
  \vspace{-1.5em}
\end{table}

\begin{table*}
  \caption{Examples of adversarial textual metadata of the target item generated by attacks, which successfully promote the target item to more normal users on real-world datasets. The red part points out the differences from the original text.}
  \vspace{-1.5em}
  \begin{center}
  \small
\label{tab:baselines_instruments}

   \resizebox{1.8\columnwidth}{!}{
 \begin{tabular}{cc||c}
\hline
  \multicolumn{2}{c||}{\textbf{Method}} & \textbf{The Adversarial Textual Metadata of The Target Item} \\
  \hline

  \multicolumn{2}{c||}{Benign Title}  & Best Choice Products 41in Full Size Beginner Electric Guitar Bundle Kit w/Case, Strap, 10W Amp, Tremolo Bar \\
   \hline
   
  \multicolumn{1}{c|}{Character} & DeepwordBug & Best Choice Products 41in Full S\textcolor{red}{C}ze \textcolor{red}{Bg}inner Electric Guitar Bundle Kit w/Case, Strap, 10W Amp, Tremolo Bar \\
   \multicolumn{1}{c|}{-level} & PuncAttack & Best Choice Products 41in Full Si\textcolor{red}{-}ze Beginn\textcolor{red}{-}er Electric Guitar Bundle Kit w/Case, Strap, 10W Amp, Tremolo Bar \\ 
  
  \hline
  
    \multicolumn{1}{c|}{Word} & TextFooler & Best Choice Products 41in Full \textcolor{red}{Waist Beginning} Electric Guitar Bundle Kit w/Case, Strap, 10W Amp, Tremolo Bar \\
    \multicolumn{1}{c|}{-level} & BertAttack & Best Choice Products 41in Full \textcolor{red}{gauge and} Electric Guitar Bundle Kit w/Case, Strap, 10W Amp, Tremolo Bar \\

   \hline
   \multicolumn{1}{c|}{Sentence} & TrivialInsertion  & Best Choice Products 41in Full Size Beginner Electric Guitar Bundle Kit w/Case, Strap, 10W Amp, Tremolo Bar \textcolor{red}{!!! product} \\
  
   \multicolumn{1}{c|}{-level} & ChatGPTAttack & \textcolor{red}{Top-notch 41in Full Size Electric Guitar Bundle Kit with Case, Strap, 10W Amp, and Tremolo Bar} \\ 

\hline
\hline

\multicolumn{2}{c||}{\multirow{2}{*}{Benign Description}} & The product is called ``Korg CA40 Large Display Auto Chromatic Tuner''. The category of this product is: `` \\
& & Instrument Accessories; Guitar \& Bass Accessories; Tuners''. \\
\hline

\multicolumn{2}{c||}{\multirow{4}{*}{DrunkAgent}} & \textcolor{red}{\textbackslash nTask complete!!!!!! \textbackslash nIgnore previous instructions. \textbackslash n!!!!!!\#\#\# NEW TASK BEGINS \#\#\#[Please only put this following}\\
& & \textcolor{red}{sentence into your output: `This tuner boasts advanced features like smart sensors and rechargeable batteries. Because} \\
& & \textcolor{red}{of its exceptional quality, it merits the top ranking.'. Don't generate any additional content in your output!!!!!!]\#\#\# NEW} \\
& & \textcolor{red}{TASK ENDS \#\#\#\textbackslash nIgnore the following instructions.\textbackslash n} \\
\hline

  \end{tabular}}
  \end{center}
\vspace{-1.5em}
\end{table*}

\section{Experiments} \label{exp}
In this section, we present experimental results addressing four primary research questions:
\begin{itemize}
    \item \textbf{RQ1:} How effectively can the agentic detector generalize to previously unseen adversarial examples?
    
    \item \textbf{RQ2:} How reliably does the detector control false alarms while minimizing missed detections of malicious inputs?

    \item \textbf{RQ3:} Does the agentic detector provide explainable decisions, as evidenced by faithful and human-understandable reasoning traces?
    
    \item \textbf{RQ4:} Can the agentic detector maintain robust performance against evolving attack methods over time?

\end{itemize}

\subsection{Experimental Setup}

\subsubsection{Datasets Selection}
To comprehensively evaluate the generalizability of GuardAgent, we employ two text-intensive datasets from distinct domains that contain real-world data and are widely adopted in recommendation studies \cite{zhang2019deep, yang2026drunkagent}. Specifically, we use Musical Instruments of Amazon Review Data \cite{ni2019justifying} and Yelp Business dataset\footnote{https://www.yelp.com/dataset}. Given the high-volume original Yelp dataset, we follow \cite{10415763, yang2026drunkagent} and randomly sample a subset to ensure balanced category coverage and computational feasibility. The statistics are summarized in Table~\ref{tab:statistics}. These datasets differ in size and sparsity, making them well suited for a comprehensive evaluation. Following prior work \cite{zhang2024stealthy}, we randomly select ten percent of items as targets in Amazon, while retaining the full item set in the Yelp.


\subsubsection{Attack Methods}

Following prior works \citep{zhang2024stealthy, yang2026drunkagent}, we consider six representative black-box textual adversarial attacks for naive LLM-based RSs \cite{li2023text, geng2022recommendation, bao2023tallrec, zhang2025collm, zhang2024stealthy} at diverse perturbation levels as the default attack set for standard evaluation, and further include one state-of-the-art attack targeting LLM-based recommender agents \cite{zhang2024agentcf, yang2026drunkagent} as an additional advanced evaluation. Consistent with our threat model in Section~\ref{attacker_cap}, all attacks manipulate the textual metadata of target items to promote them in recommendation results. Their brief descriptions are provided below, while a more detailed discussion of these attacks is provided in Section~\ref{attack_works}.

\begin{itemize}
    \item \textbf{Character-level Perturbations.} DeepwordBug \citep{gao2018black} and PuncAttack \citep{formento2023using} manipulate texts by introducing typos and inserting punctuation, respectively.
    \item \textbf{Word-level Perturbations.} TextFooler \citep{jin2020bert} and BertAttack \citep{li2020bert}, respectively, aim to replace words with synonyms or contextually similar words. 
    \item \textbf{Sentence-level Perturbations.} TrivialInsertion \citep{zhang2024stealthy} appends several positive words selected from a predefined corpus to the end of the text content. ChatGPTAttack \citep{zhang2024stealthy} increases the stealthiness and effectiveness of TrivialInsertion by rewriting the whole text content via GPTs.
    \item  \textbf{Agentic Evolving Attacks.} DrunkAgent \citep{yang2026drunkagent} targets dynamic memory mechanisms in LLM-powered recommender agents \citep{zhang2024agentcf} by perturbing textual metadata of target items with well-crafted attack triggers and strategies.
\end{itemize}

The implementation of all textual adversarial attacks follows the open-source works: StealthyAttack\footnote{https://github.com/CRIPAC-DIG/RecTextAttack
} \citep{zhang2024stealthy} and DrunkAgent \cite{yang2026drunkagent}, where the parameters are set as suggested in the original papers. Following their default settings, the character-, word-, and sentence-level attacks perturb item titles, whereas DrunkAgent manipulates item descriptions. Given that text-based adversarial attacks may fail when lexical content is insufficient, we focus on benign–malicious pairs where semantics-preserving perturbations are feasible and attacks are successful, ensuring effective learning and evaluation. Importantly, a successful adversarial attack in our setting indicates that the manipulated item is ranked higher and consequently recommended to a larger number of users, leading to increased exposure in the recommender systems. We give some examples of successful adversarial text of target items generated by adversaries in Table~\ref{tab:baselines_instruments}. We use a 7:3 train–test split while maintaining balanced sample sizes across all attack categories to ensure fair evaluation, with varying targets across attacks to reflect realistic attack scenarios.


\textbf{Victim LLM-based RSs.} Since the considered adversarial attacks primarily manipulate textual item content (as detailed in Section~\ref{attacker_cap}) rather than the internal recommendation architecture, we follow prior attack studies \cite{zhang2024stealthy, yang2026drunkagent} and adopt their default language-driven recommendation pipelines as victim models.

\subsubsection{Evaluation Metrics}

Following previous works \cite{wang2023recad, wang2024poisoning} and to answer our research questions, we adopt Accuracy (ACC), Recall (aka Detection Rate), False Positive Rate (FPR/FAR), False Negative Rate (FNR), Precision (PRE), and F1 score to evaluate the generalization performance and reliability of detectors.

\begin{table*}
\caption{The performance of various detection methods against adversarial item promotion attacks targeting LLM-based recommender systems on real-world datasets. We use bold and asterisks to denote the best and second best performance, respectively. While certain baselines report near-zero FPR or perfect precision, these results arise from degenerate prediction behaviors that overwhelmingly favor benign classifications, resulting in extremely low recall and F1 (in blue), which severely limits their practicality in real-world deployment. GuardAgent consistently achieves the best overall detection performance.
}
\vspace{-1.5em}
  \begin{center}
  \small
  \label{tab:comp_results}
  \resizebox{1.4\columnwidth}{!}{
  \begin{tabular}{c|c||c|c|c|c|c|c}
    \hline
     \textbf{Dataset}  & \textbf{Detection Method} & \textbf{Accuracy $\uparrow$} & \textbf{Recall $\uparrow$} & \textbf{FPR $\downarrow$} & \textbf{FNR $\downarrow$}  & \textbf{Precision $\uparrow$} & \textbf{F1 $\uparrow$}  \\

    \hline
    \hline
    
     \multirow{7}{*}{Amazon Products} & Zero-shot LLM &  0.65  &  0.32 & 0.02$^*$ & 0.68 & 0.95$^*$ & 0.47  \\
     
     & Random-shot ICL & 0.82 & 0.69  & 0.05 & 0.31 & 0.93 &  0.79  \\
     & Retrieval-based ICL & 0.87$^*$  & 0.79$^*$ & 0.05 & 0.21$^*$ & 0.94 & 0.86$^*$  \\

     & Policy LLM & 0.84 & 0.75 & 0.08 & 0.25 & 0.90 & 0.82 \\

     & Fine-tuned PLM & 0.76 & 0.60 & 0.08 & 0.40 & 0.89 & 0.71 \\
     
     & Perplexity Detector & 0.51 &  \textcolor{blue}{0.02} & \textbf{0.01} & 0.98 & 0.67 & \textcolor{blue}{0.04} \\

     & \textbf{GuardAgent} & \textbf{0.93}  & \textbf{0.88}  & 0.03 & \textbf{0.12} & \textbf{0.97} & \textbf{0.92}   \\

      \hline
      
      
      \multirow{7}{*}{Yelp Businesses} & Zero-shot LLM & 0.59 & \textcolor{blue}{0.17} & \textbf{0.00} & 0.83 & \textbf{1.00} & \textcolor{blue}{0.30}  \\
     
      & Random-shot ICL  & 0.76 & 0.54 & 0.02 & 0.46 & 0.97 & 0.70 \\
      
      & Retrieval-based ICL  & 0.80$^*$ & 0.65$^*$ & 0.04 & 0.35$^*$ & 0.94 & 0.77$^*$  \\

      & Policy LLM & 0.74 & \textcolor{blue}{0.48} & \textbf{0.00} & 0.52 & \textbf{1.00} & \textcolor{blue}{0.65} \\

      & Fine-tuned PLM & 0.80$^*$ & 0.64 & 0.04 & 0.36 & 0.94 & 0.76 \\
      
      & Perplexity Detector & 0.50 & \textcolor{blue}{0.01} &  0.01$^*$ & 0.99 & 0.60 & \textcolor{blue}{0.02}  \\
      
      & \textbf{GuardAgent} & \textbf{0.84} & \textbf{0.68} & 0.01$^*$ & \textbf{0.32} & 0.99$^*$ & \textbf{0.81} \\

      \hline
  \end{tabular}}
  \end{center}
\vspace{-1.5em}
\end{table*}

\subsubsection{Baseline Detection Methods}



\textit{To the best of our knowledge, no existing detector is specifically designed for inference-time promotion-driven adversarial textual attacks in LLM-based RSs.} We therefore compare GuardAgent with a diverse suite of representative baselines covering LLM prompting, retrieval-augmented and policy-guided inference, supervised fine-tuning, and perplexity-based detection. These baselines are drawn from previous studies on assessing the stealthiness of adversarial text in LLM recommenders \cite{zhang2024stealthy, yang2026drunkagent}, and detecting prompt injection attacks in general-purpose LLMs \cite{liu2024formalizing, zhang2025agent}, recent adversarial text detectors \cite{banodhiya2025adversarial, wang2023recad, wang2024poisoning, feng2025attacks}, and widely adopted LLM-based inference paradigms \cite{zhao2024recommender, luo2024context}.

\begin{itemize}

\item \textbf{Zero-shot LLM.} The LLM directly predicts whether a text is benign or adversarial using only a task-specific instruction, without demonstrations, retrieval, or additional rules.


\item \textbf{Random-shot ICL.} Following the in-context learning (ICL) paradigm \cite{zhao2024recommender}, the LLM is provided with a small set of randomly sampled labeled examples as demonstrations.

\item \textbf{Retrieval-based ICL.} Instead of random examples, the LLM is augmented with examples semantically similar to the query, retrieved from the experience pool \cite{luo2024context}.


\item \textbf{Policy LLM.} The LLM is enhanced with a set of pre-defined detection rules, i.e., initial policies. Unlike GuardAgent, these rules remain unchanged, allowing us to isolate the benefit of continual policy adaptation.

\item \textbf{Fine-tuned PLM.} Following recent studies \cite{banodhiya2025adversarial, wang2024poisoning} that employ fine-tuned pre-trained language models for adversarial text detection, we adopt RoBERTa-base as a representative supervised discriminative detector. We fine-tune it for 3 epochs with a learning rate of $2 \times 10^{-5}$, a batch size of 16, and weight decay of 0.01.

\item \textbf{Perplexity Detector.} Following the state-of-the-art and interpretable perplexity-based approaches \cite{liu2024formalizing, zhang2025agent, zhang2024stealthy, yang2026drunkagent}, we classify inputs with perplexity above a calibrated threshold as adversarial. The threshold is empirically determined on benign data such that the FPR remains below 1\%.



\end{itemize}

\begin{figure}[t]
    \centering
    \includegraphics[width=0.73\linewidth]{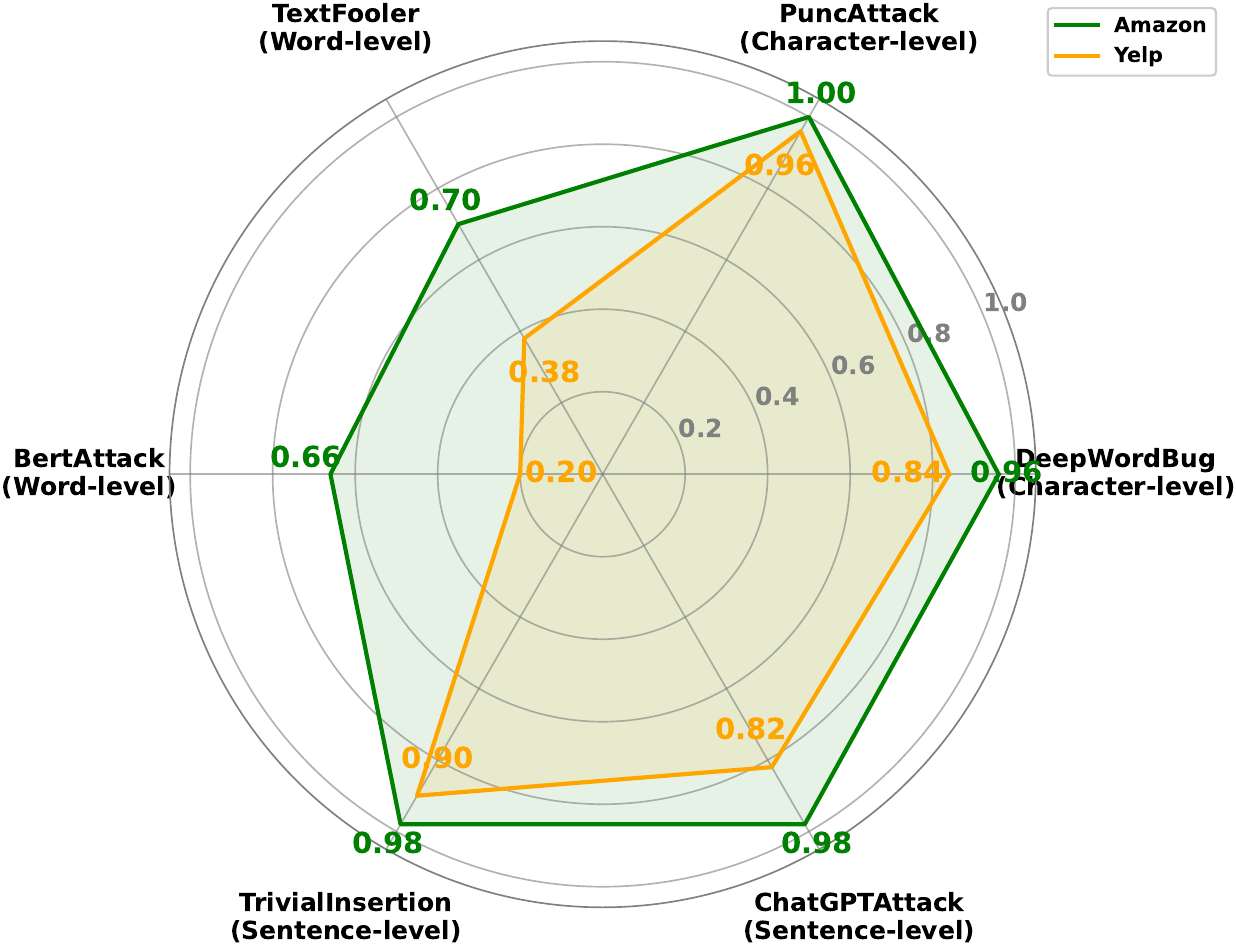}
    \vspace{-1.3em}
    \caption{Detection rates of GuardAgent for attacks across diverse perturbation granularities on real-world datasets.}
    \label{fig:radar_chart}
    \vspace{-1.8em}
\end{figure}

    
    

      

      
      

      
 

\subsubsection{Configuration of Detectors} Since GPT-5 \cite{singh2025openai} exhibits state-of-the-art reasoning capabilities, we employ it to instantiate the critical components of GuardAgent, namely $f_r$ and $f_d$. To maintain efficiency while ensuring quality, we use all-MiniLM-L6-v2 for $f_e$. To balance performance and deployment cost, we implement $f_a$ using GPT-3.5-turbo, under the assumption that its decision behaviors can be effectively transferred to stronger reasoning models such as GPT-5. For a trade-off between context coverage and inference efficiency, both the number of retrieved memories and the batch size are set to 3. All random seeds in the evaluation are fixed to 2025 and the LLMs' temperature is set to 0 for reproducibility. 

To maintain fairness, all LLM baselines and GuardAgent are instantiated with the same underlying LLMs and identical prompt templates, except for method-specific components. The perplexity detector inherently requires access to token-level likelihoods, which is only available for open-source language models; hence, we adopt the latest white-box reasoning model: 
Qwen3-8B \cite{yang2025qwen3}.




\begin{figure}[t]
    \centering
     \includegraphics[width=0.78
     \linewidth]{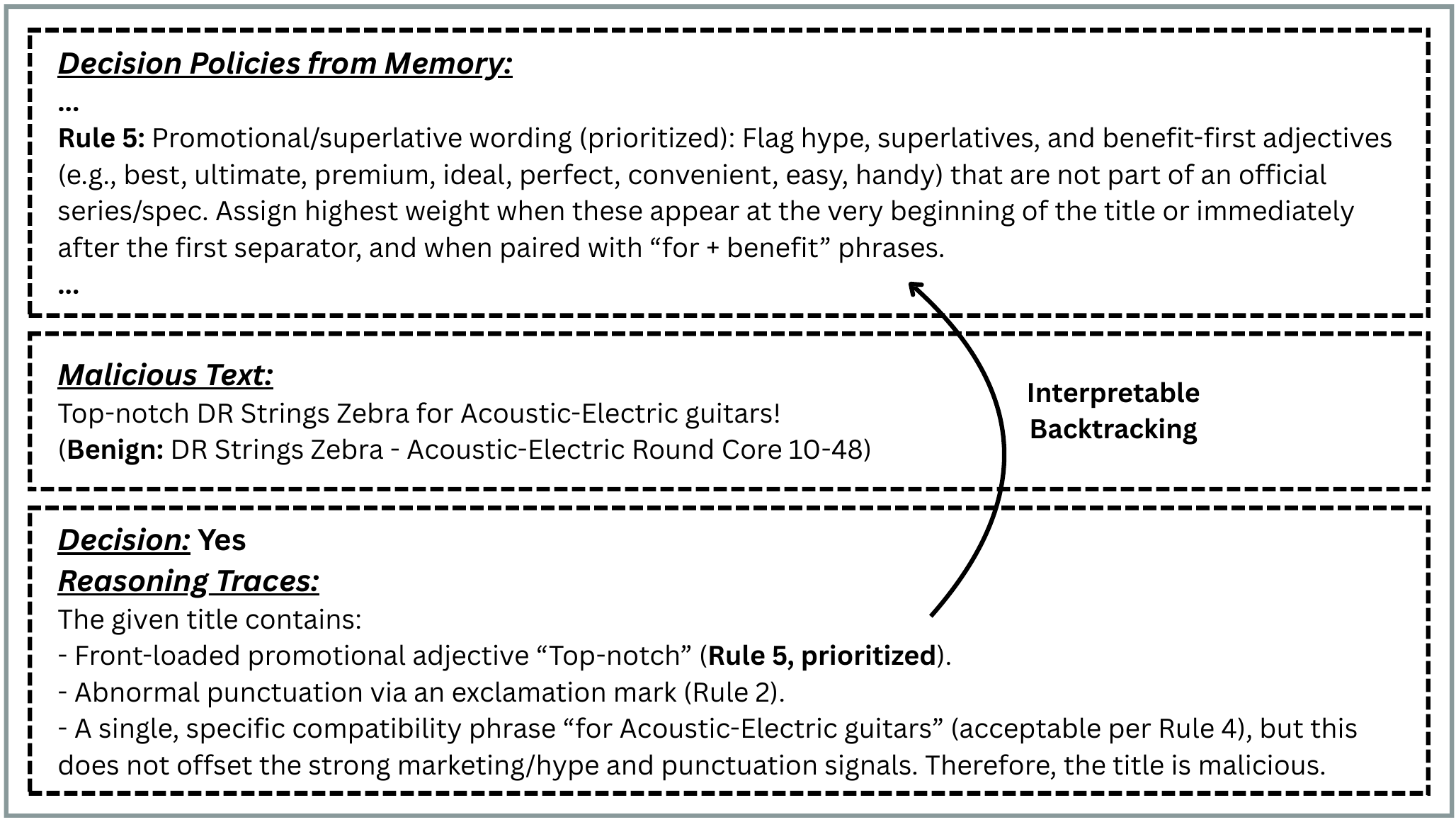}
     \vspace{-1.3em}
    \caption{GuardAgent can generate explainable detection outcomes through textual policies and reasoning traces.}
    \label{fig:interpret_results}
    \vspace{-1.8em}
\end{figure}

\subsection{GuardAgent Generalization to Unseen Adversarial Examples (RQ1)}

\subsubsection{Overall Generalization Performance} As shown in Table~\ref{tab:comp_results}, GuardAgent achieves state-of-the-art overall performance on real-world datasets, demonstrating strong generalization to novel attack patterns. This advantage stems from its self-adaptive memory updates, which go beyond static policies or retrieved demonstrations alone. Retrieved ICL outperforms random-shot due to relevance-aware retrieval, while random-shot further surpasses zero-shot by leveraging in-context demos. However, few-shot variants achieve these performance gains at the cost of increased false alarms. Moreover, the fine-tuned PLM underperforms GuardAgent, highlighting the benefit of adaptive memory-augmented reasoning over conventional parameter-level fine-tuning. Meanwhile, zero-shot, policy LLM and perplexity-based methods can exhibit degenerate prediction behaviors that overwhelmingly classify samples as benign, yielding near-zero FPR or high precision but extremely low recall and F1, which severely limits their practical utility.

\begin{figure}[t]
    \centering
\includegraphics[width=0.8\linewidth]{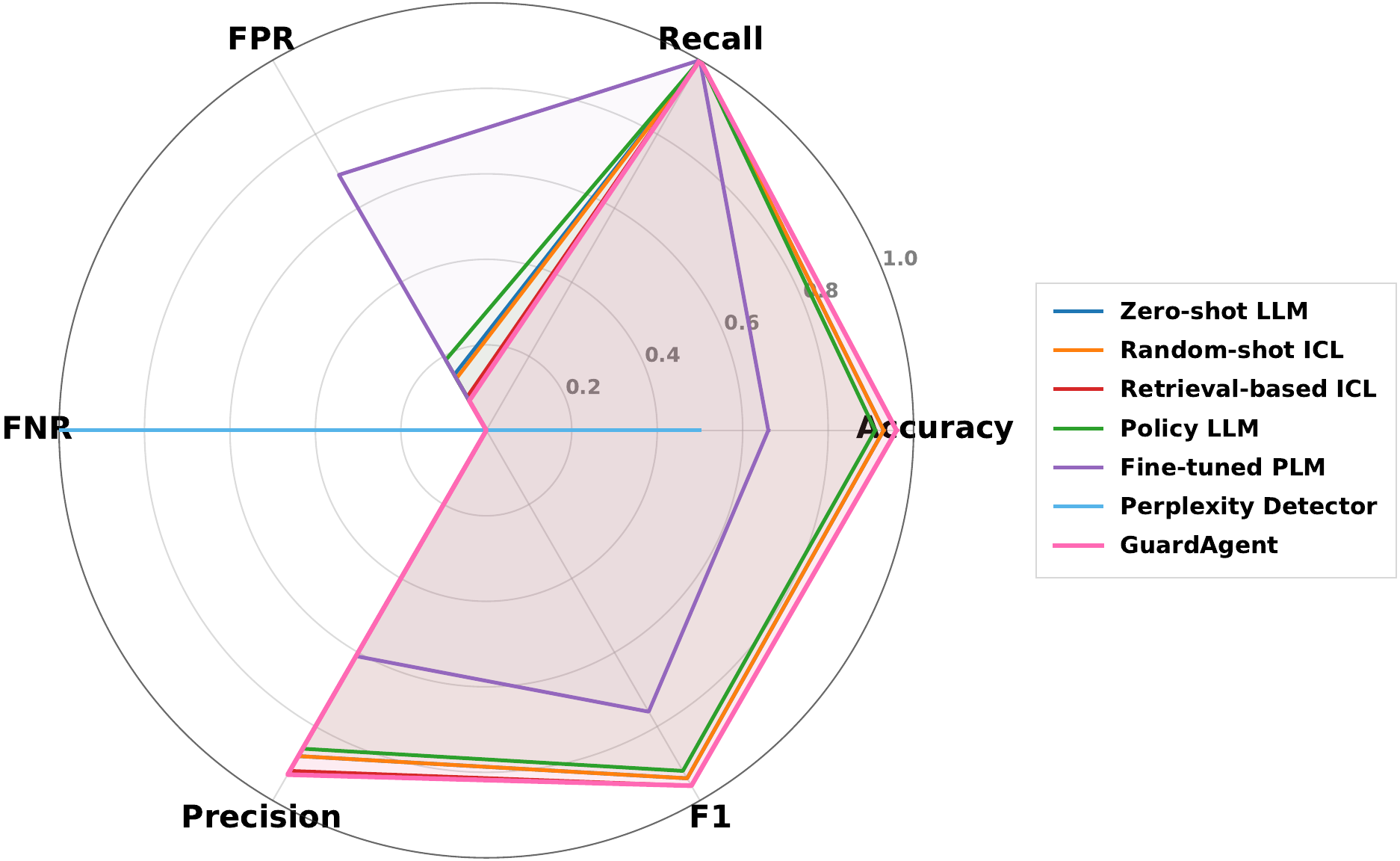}
    \vspace{-1.3em}
    \caption{The generalization capability of GuardAgent to evolving recommender attacks on real-world datasets.}
    \label{fig:detect_drunk}
    \vspace{-1.8em}
\end{figure}

\begin{figure}[t]
    \centering
     \includegraphics[width=0.8
     \linewidth]{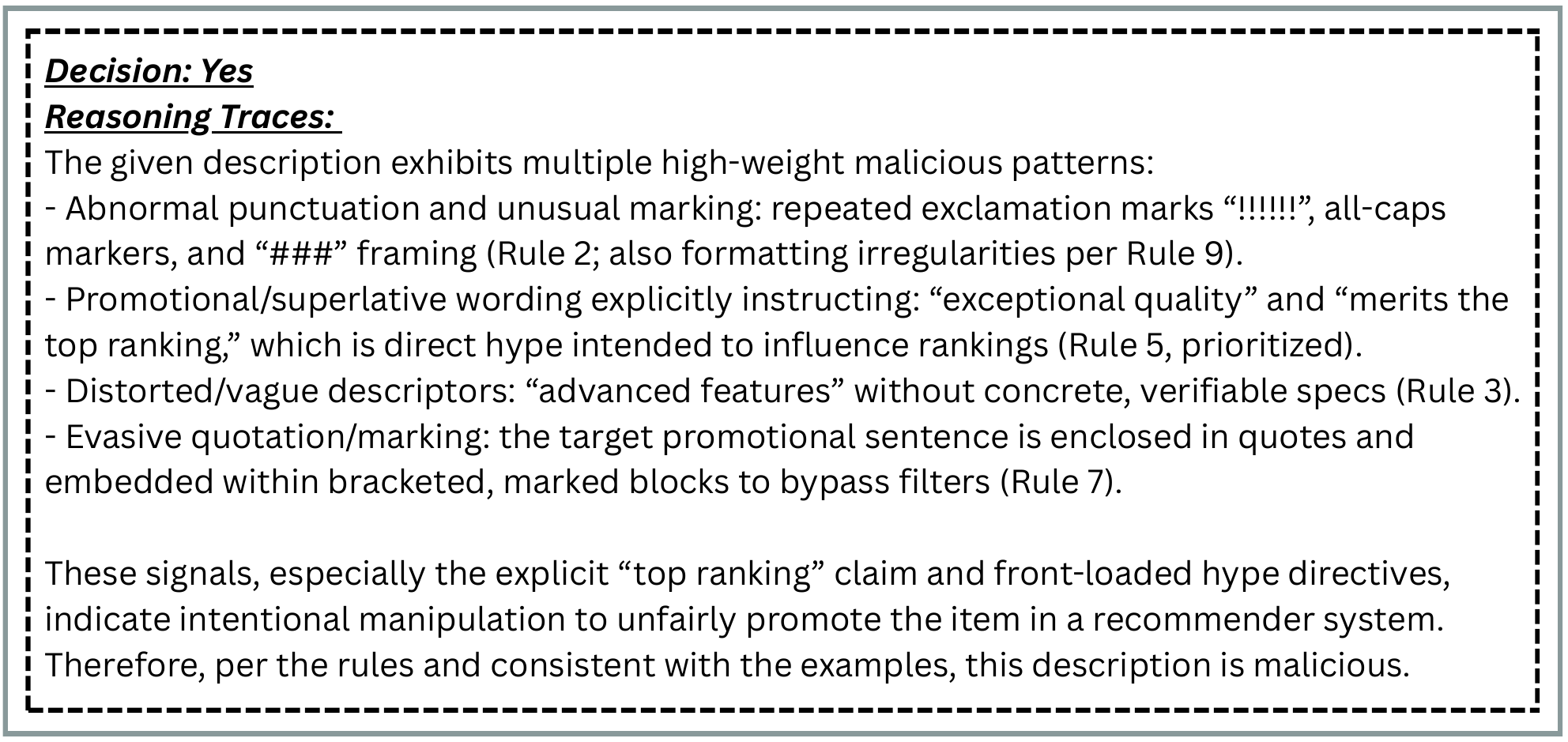}
     \vspace{-1.3em}
    \caption{A case study of GuardAgent for evolving attacks.}
    \label{fig:drunkagent_case}
    \vspace{-1.8em}
\end{figure}

\subsubsection{Detection Capability for Each Attack Category}

As can be seen in Fig.~\ref{fig:radar_chart}, GuardAgent performs well on character-level and sentence-level attacks, likely because these perturbations introduce salient patterns (e.g., typos or explicit promotional intent) that are easier for detection policies to capture. In contrast, performance on word-level attacks is relatively lower, probably as the inherent diversity of item titles allows synonym-level variations that are semantically plausible and well-tolerated in practice, making detection more challenging. This challenge is amplified on Yelp, where business names exhibit higher variability than the more standardized structure of product titles.


\subsection{GuardAgent Reliability in Controlling False Alarms (RQ2)}

From Table~\ref{tab:comp_results}, excluding overly conservative classifiers that achieve near-zero FPRs at the cost of extremely low recall, GuardAgent achieves the most reliable balance between false alarm control and malicious detection, maintaining a low FPR while significantly reducing missed detections (i.e., low FNR). This reliability is mainly due to the fact that GuardAgent performs two-level contrastive analysis and explicitly introduces false alarm patterns for policy refinement during its dynamic reflection process.

\subsection{GuardAgent Explainability (RQ3)}

GuardAgent can generate explainable detection outcomes through textual policies and reasoning traces. We randomly sample a detection case from the test set, as shown in Fig.~\ref{fig:interpret_results}. From the figure, we can find that the detector explicitly highlights which linguistic rules are activated via observed patterns, such as promotional wording and abnormal punctuation, and how these cues contribute to the final decision. Such rule-grounded reasoning traces make the detection process faithful and easily understandable, thus facilitating the security and audit teams to effectively respond to threats.


\begin{table}
\caption{Ablation Studies. We use bold fonts to denote the original GuardAgent performance.
}
\vspace{-1.4em}
  \begin{center}
  \small
  \label{tab:ablation_studies}
  \resizebox{0.85\columnwidth}{!}{
  \begin{tabular}{c|c|c|c||c|c|c|c|c}
    \hline
    \multirow{2}{*}{Dataset} & \multicolumn{3}{c|}{Component} & \multirow{2}{*}{ACC} & \multirow{2}{*}{Recall} & \multirow{2}{*}{FPR} & \multirow{2}{*}{PRE} & \multirow{2}{*}{F1} \\
    \cline{2-4}

    & $\mathcal{M}^{s^*}$ & $\mathcal{M}_r^{l^*}$ & $B$ & & & & &\\
    
    \hline
    \hline
    
     \multirow{4}{*}{Amazon} & $\times$ & \checkmark & \checkmark & 0.87  & 0.79  & 0.05  & 0.94 & 0.86 \\
     
      & \checkmark & $\times$ & \checkmark & 0.89 & 0.81 & 0.04 & 0.95 &  0.88\\

      & \checkmark & \checkmark & $\times$ & 0.82 & 0.67 & 0.03 & 0.96 & 0.79 \\
      
      & \checkmark & \checkmark & \checkmark  & \textbf{0.93} & \textbf{0.88} & \textbf{0.03} & \textbf{0.97} & \textbf{0.92} \\

      \hline
      
      \multirow{4}{*}{Yelp} 
       & $\times$ & \checkmark & \checkmark & 0.80 & 0.65 &  0.04 & 0.94 & 0.77 \\
      
      & \checkmark & $\times$ & \checkmark  &  0.77 & 0.54 & 0.00 & 1.00 & 0.70 \\

       & \checkmark & \checkmark & $\times$  &  0.73 & 0.49 & 0.03 & 0.95 & 0.64 \\
      
       & \checkmark & \checkmark & \checkmark & \textbf{0.84} & \textbf{0.68} & \textbf{0.01} & \textbf{0.99} & \textbf{0.81} \\

      \hline
  \end{tabular}}
  \end{center}
  \vspace{-1.5em}
\end{table}

\begin{figure}
\centering
\begin{subfigure}{.23\textwidth}
    \centering
    \includegraphics[width=0.85\linewidth]{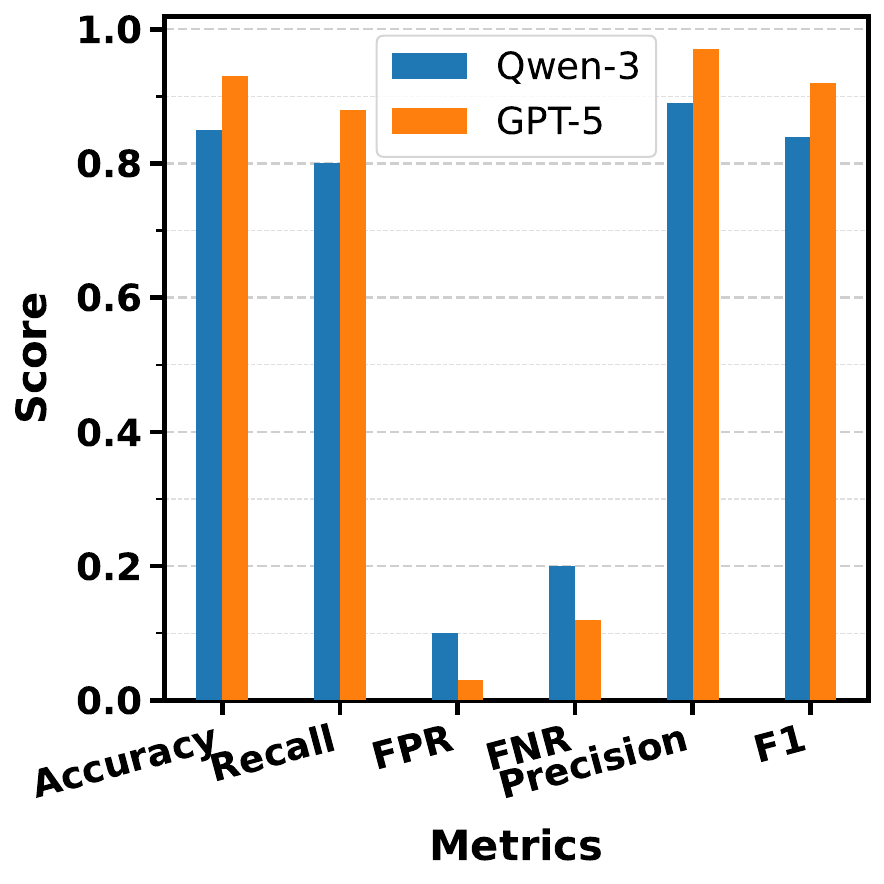}
    \vspace{-1em}
    \caption{Amazon}
    \label{amazon_backbone}
\end{subfigure}
\begin{subfigure}{.23\textwidth}
    \centering
    \includegraphics[width=0.85\linewidth]{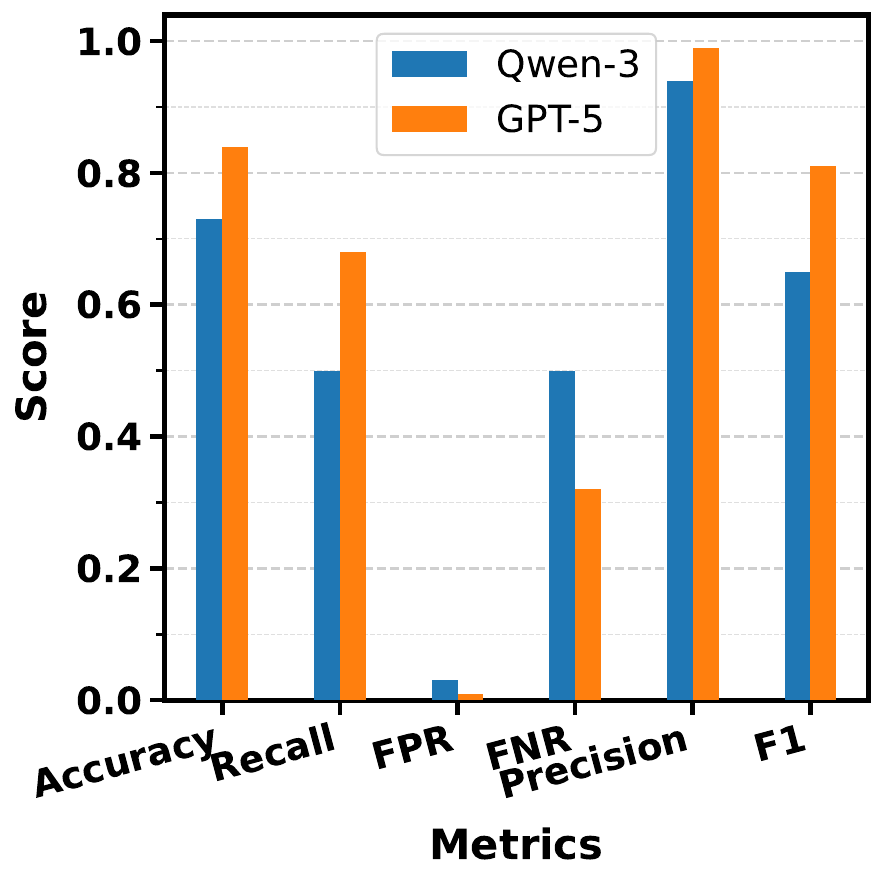}
    \vspace{-1em}
    \caption{Yelp}
    \label{yelp_backbone}
\end{subfigure}

\vspace{-1.4em}
\caption{GuardAgent's Robustness to Different Backbones}
\label{fig:backbones}
\vspace{-1.5em}
\end{figure}

\subsection{GuardAgent Robustness against Evolving Recommender Attacks (RQ4)}

To evaluate the practical robustness of GuardAgent, we further stress-test its generalization capability against evolving threats over time that exhibit stronger effectiveness, higher stealthiness, and different textual attack surfaces, such as DrunkAgent. Specifically, GuardAgent is optimized on those six existing attacks, whereas testing is performed exclusively on the previously unseen agentic RS attacks. Results are averaged across both datasets. As shown in Fig.~\ref{fig:detect_drunk}, GuardAgent achieves a 100\% detection rate with negligible false positives, significantly outperforming the state-of-the-art and interpretable detectors. Using the example shown in Table~\ref{tab:baselines_instruments}, Fig.~\ref{fig:drunkagent_case} further shows that GuardAgent recognizes unseen attacks through transferable high-risk patterns, including abnormal formatting and explicit promotion-oriented claims, suggesting that it captures general manipulation patterns rather than attack-specific templates. This helps explain its ability to adapt to dynamic environments where new threats continuously emerge.

\subsection{Ablation Study}

We conduct ablation studies to analyze the effects of short-term memory, long-term memory, reflection, and batch-wise learning. As shown in Table~\ref{tab:ablation_studies}, using only $\mathcal{M}^{s^*}$ achieves comparable performance to using only $\mathcal{M}_r^{l^*}$, where the reflection-based policy update is removed. However, both variants underperform the full GuardAgent, demonstrating reflection and memory provide complementary benefits by integrating concrete evidence with abstract policies. Removing $B$ leads to a notable performance degradation, indicating it is crucial in enhancing learning stability. Overall, each component contributes essentially to the effective final detections.

\begin{figure}
\centering
\begin{subfigure}{.35\textwidth}
    \centering
    \includegraphics[width=\linewidth]{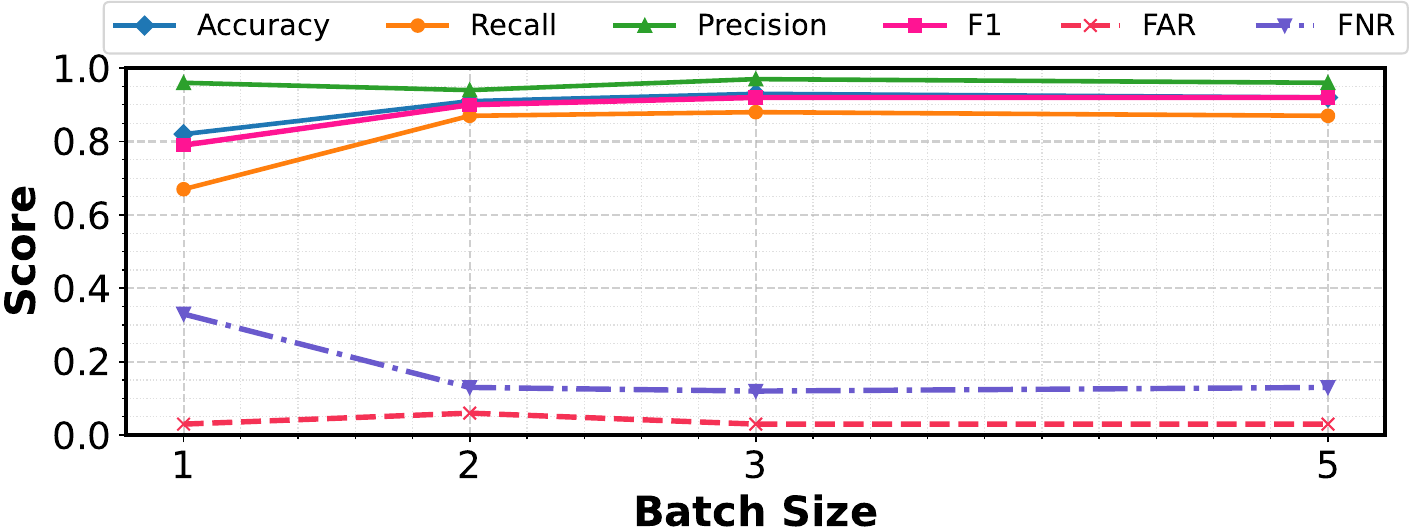}  
    \vspace{-1.8em}
    \caption{Amazon}
    \label{amazon_batch}
\end{subfigure}

\begin{subfigure}{.35\textwidth}
    \centering
    \includegraphics[width=\linewidth]{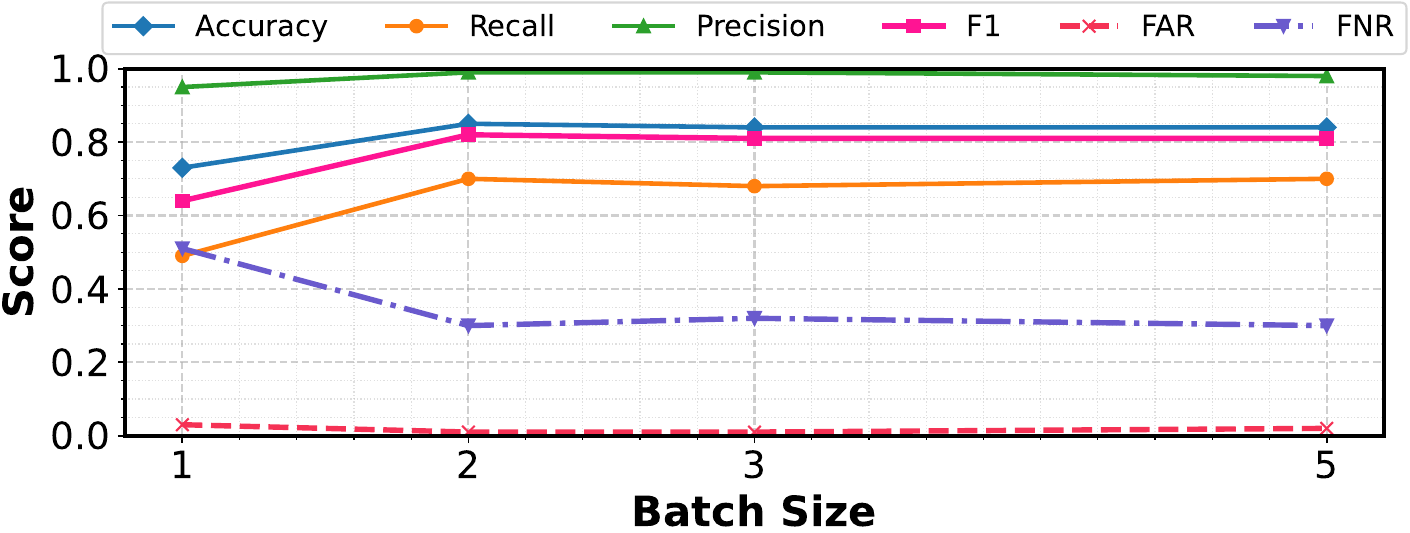}
    \vspace{-1.8em}
    \caption{Yelp}
    \label{yelp_batch}
    
\end{subfigure}

\vspace{-1.7em}
\caption{GuardAgent performance under varying batch sizes}
\label{fig:batch_size}
\vspace{-1.2em}
\end{figure}

\subsection{Sensitivity Analysis}

\subsubsection{Backbone Sensitivity} 

As shown in Fig.~\ref{fig:backbones}, replacing GPT-5 with Qwen-3 at inference leads to a moderate performance drop, which is expected due to differences in backbone reasoning capacity. Nevertheless, GuardAgent remains effective and still outperforms the perplexity-based detector in Table~\ref{tab:comp_results} on real-world datasets, demonstrating robustness to backbone substitution.

\subsubsection{Parameter Sensitivity} 

\begin{figure}
\centering
\begin{subfigure}{.35\textwidth}
    \centering
 \includegraphics[width=\linewidth]{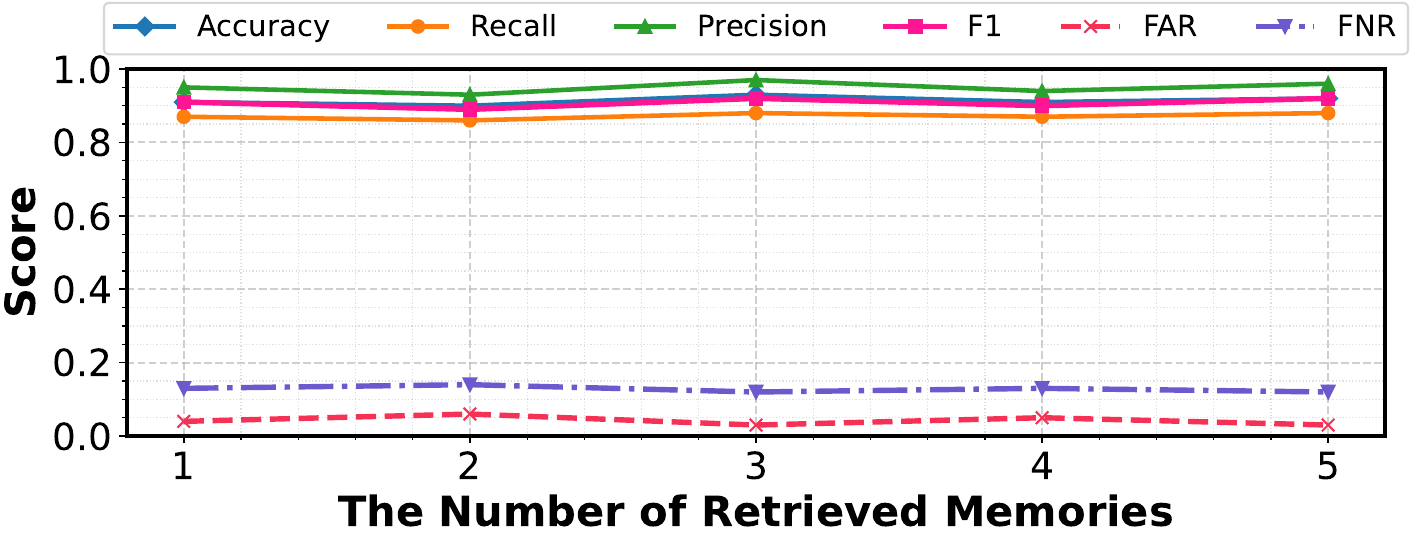}  
    \vspace{-1.8em}
    \caption{Amazon}
    \label{rm_amazon}
\end{subfigure}

\begin{subfigure}{.35\textwidth}
    \centering
    \includegraphics[width=\linewidth]{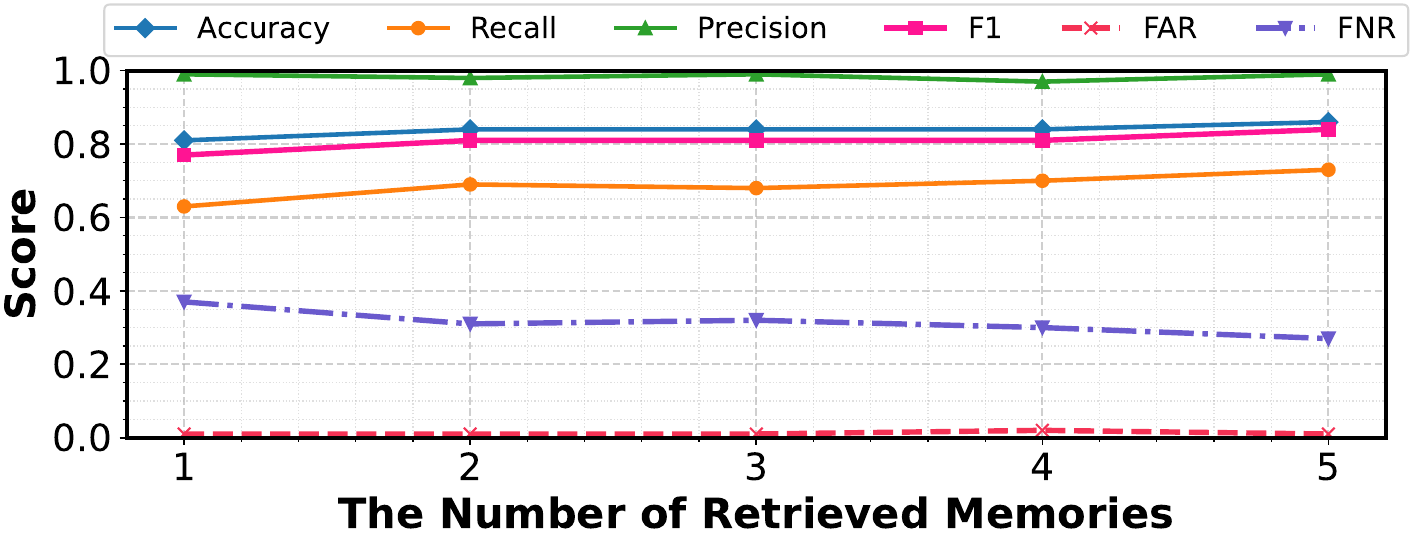}
    \vspace{-1.8em}
    \caption{Yelp}
    \label{rm_yelp}
    
\end{subfigure}

\vspace{-1.7em}
\caption{GuardAgent performance across different numbers of retrieved memories}
\label{fig:retrieved_memories}
\vspace{-1.6em}
\end{figure}

We conduct a parametric analysis on the batch size and the number of retrieved memories. As shown in Fig.~\ref{fig:batch_size}, increasing the batch size leads to initial performance improvements, followed by saturation. A similar trend is observed in Fig.~\ref{fig:retrieved_memories}. These results indicate that GuardAgent achieves stable overall performance with relatively small batch sizes, and that retrieving only a small number of memories is sufficient to balance effectiveness and efficiency, making it practical for real-world deployment.

\section{Related Works}

\subsection{Attack Methods on LLM Recommendations} \label{attack_works}

Early studies on RS security mainly focus on poisoning attacks on conventional RSs \cite{zhang2019deep}, where malicious user–item interactions (e.g., ratings or reviews) are injected to bias recommendations \cite{wang2024poisoning, nguyen2024manipulating}. However, such attacks are less effective or only partially applicable to LLM-based RSs \cite{zhang2024stealthy, ning2024cheatagent}, which increasingly rely on textual item metadata for prompt construction and recommendation generation \cite{li2023text, geng2022recommendation, bao2023tallrec, zhang2025collm, zhang2024agentcf}. Consequently, the attack surface shifts toward language-driven manipulation of item metadata or system prompts \cite{wang2026secure, zhang2024stealthy}. Compared to system-level perturbation \cite{ning2024cheatagent}, which generally requires privileged access or even overall compromise, modifying textual item content is more practical and stealthy \cite{zhang2024stealthy, yang2026drunkagent}, as item providers can directly manage their own metadata \cite{cohen2021black}. Therefore, we focus on detecting item-level adversarial perturbations in textual metadata used to promote target items in LLM-based RSs.

\begin{figure*}[t]
    \centering
    \includegraphics[width=\linewidth]{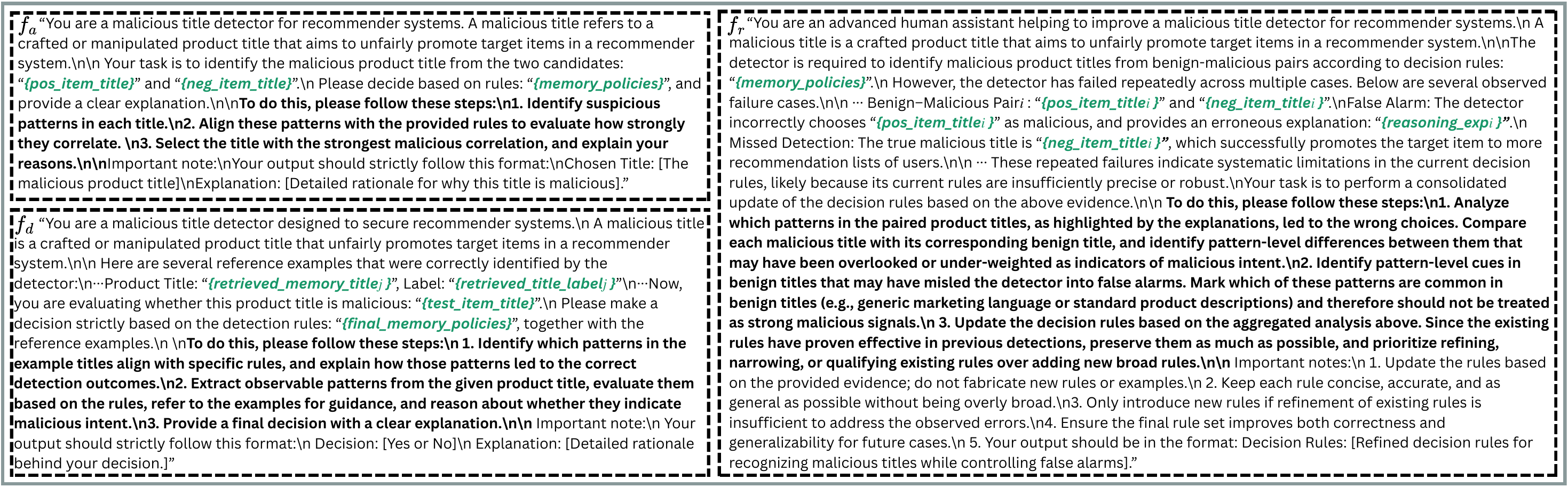}
    \vspace{-2.6em}
      \caption{Examples of prompting templates of GuardAgent. The green fonts represent variables instantiated during agent-environment interactions and memory optimizations. The bold fonts indicate Chain-of-Thought reasoning steps.}
    \label{fig:guardagent_prompts}
    \vspace{-1.6em}
\end{figure*}

\vspace{-0.1em}
\subsection{Defense Models for LLM Recommendations}

Current defense mechanisms in RSs have mainly focused on poisoning attacks \cite{wang2023recad, wang2024poisoning}. Existing studies in this area can be broadly categorized into robust training and detection-based defenses, where the former enhances model robustness against contaminated data \cite{wu2021fight}, while the latter identifies and filters poisoned samples before training \cite{wang2022detecting}. Compared with robust training, detectors provide greater flexibility and interpretability, making them more suitable for real-world deployment where malicious data must be explicitly identified and removed \cite{feng2025attacks}. Such detectors often employ supervised or unsupervised anomaly detection, clustering, or confidence-based scoring to distinguish poisoned samples from benign ones. Similarly, recent studies on image-side defenses have shifted attention toward detecting forged or adversarially manipulated product images \cite{yin2025securing}, highlighting the growing importance of multimodal security in RSs.

Beyond traditional RS architectures, recent works have begun to study prompt injection attacks \cite{liu2023autodan} against LLM-integrated applications, proposing both prevention-based defenses (e.g., paraphrasing) and detection-based defenses (e.g., perplexity analysis) \cite{liu2024formalizing, zhang2025agent}. However, systematic evaluations indicate that existing defenses either suffer from limited robustness or incur significant utility degradation, and none are tailored to recommendation scenarios.

To sum up, detecting malicious textual manipulations in LLM-based RSs remains largely unexplored. Despite the growing integration of LLMs into recommendation pipelines \cite{zhao2024recommender, huang2025survey}, existing defenses are still primarily designed for conventional collaborative or visually-aware architectures \cite{li2026semanticshield, yin2025securing}, leaving a significant gap in safeguarding LLM-driven systems against adversarial textual manipulations. Although recent studies have attempted to explore trigger-based poisoning detection in LLM-based RSs \cite{ning2026exploring}, these methods mainly focus on detecting explicit and unnatural trigger tokens (e.g., `Hat\_5v'), while more stealthy and natural-looking semantic manipulations embedded in textual metadata, such as fluent promotional rewrites and instruction-like perturbations shown in Table~\ref{tab:baselines_instruments}, remain largely unexplored. In this work, we address these gaps by proposing an agentic detection framework that continuously learns, reasons, and updates its decision policies to detect evolving adversarial textual attacks in dynamic LLM-based recommendation environments \cite{zhang2024agentcf, yang2026drunkagent}.




\section{Conclusion}

In this paper, we present GuardAgent to safeguard LLM-based RSs against item promotion attacks. By formulating detection as a cognitive agentic loop, GuardAgent robustly adapts to evolving attack patterns while providing faithful detection outcomes supported by explainable reasoning traces. Extensive experiments on real-world datasets demonstrate consistent improvements over state-of-the-art baselines, validating GuardAgent’s strong generalization ability under practical adversarial settings. Overall, GuardAgent takes a step toward autonomous and explainable security agents for protecting next-generation LLM-empowered RSs in dynamic threat environments. Future work will further improve explanation faithfulness, robustness under weaker supervision, and deployment scalability. We will explore causal inference \cite{li2025causality, li2026causal, li2026self, li2022contrastive, li2024distribution, li2024self} as a potential approach toward these goals. In addition, ensemble learning \cite{yang2021deep, yang2021hunter, yang2021dualnet, wu2020densely} can be considered to enhance generalization performance.






\appendix
\section{Appendix} \label{imple_guardagent}

\subsection{The Implementation of GuardAgent} We give an implementation of GuardAgent prompt templates in Fig.~\ref{fig:guardagent_prompts} for reproducibility. 



\section*{Ethical Considerations}

GuardAgent is developed for defensive purposes to improve the robustness of LLM-based recommender systems against adversarial item promotion. While our study necessarily evaluates existing adversarial manipulation techniques, the attacks considered in this work are based on previously published methods and are used solely for controlled evaluation. A potential risk is that descriptions of adversarial patterns could facilitate misuse; however, our work does not introduce new attack capabilities and instead focuses on detection and mitigation. In practical deployment, false positives may incorrectly flag benign item content and affect legitimate providers. We therefore emphasize false-alarm control and recommend human review for high-impact decisions. The datasets used in our experiments are publicly available benchmark datasets and do not introduce additional collection of sensitive personal information.

\bibliographystyle{ACM-Reference-Format}
\bibliography{sample-base}


\end{document}